\newcommand{\bra}[1]{\langle #1 \lvert}
\newcommand{\ket}[1]{\lvert #1\rangle}
\newcommand{\braket}[2]{\langle #1|#2\rangle}

\newcommand{\Tr}{\operatorname{Tr}}
\newcommand{\Jx}{J_{ex}}

\def\Q{{\@QC Q}}
\def\C{{\@QC C}}
\def\@QC#1{\mathpalette{\setbox0=\hbox\bgroup$\rm}%
  {\egroup C$\egroup\rm\rlap{\kern0.4\wd0\vrule
  width 0.05\wd0 height 0.97\ht0 depth -0.01\ht0}%
  #1\bgroup}}

\documentclass[aps,twocolumn,raggedbottom,prb,showpacs,nobalancelastpage,amssymb,groupedaddress]{revtex4}
\usepackage{graphicx}
\usepackage{amsmath}
\usepackage{amssymb}
\include{nem}

\begin{document}
\title{Hyperfine induced electron spin and entanglement dynamics in double quantum dots: The case of 
separate baths}
\author{B.\, Erbe and J.\, Schliemann}
\affiliation{Institut f\"{u}r Theoretische Physik, Universit\"at
Regensburg, 93053 Regensburg, Germany}
\date{\today}

\begin{abstract}
We consider a system of two strongly coupled electron spins in zero magnetic field, each of which is 
interacting with an individual bath of nuclear spins via the hyperfine interaction.
Applying the long spin approximation (LSA) introduced in Ref. \cite{BJEPL} (here each bath is replaced
by a single long spin), we numerically study the electron spin and entanglement dynamics. We demonstrate that the
decoherence time is scaling with the bath size according to a power law. As expected, the decaying part
of the dynamics decreases with increasing bath polarization. However, surprisingly it turns out that, under certain circumstances,
combining quantum dots of different geometry to the double dot setup has a very similar effect on the magnitude
of the spin decay. Finally, we show that even for a comparatively weak exchange coupling the electron spins
can be fully entangled.
\end{abstract}
 \maketitle

\section{Introduction}

The Loss-DiVincenco proposal is one of the most promising concepts for solid state
quantum information processing. Here electron spins confined
in semiconductor quantum dots are utilized as qubits. \cite{LossDi98, Hanson07}  
The central drawback of this approach is the fast decoherence caused by the coupling of the electron spin qubits
to the nuclear spins of the host material via the hyperfine interaction \cite{KhaLossGla02, KhaLossGla03, expMarcus, Koppens05, Petta05, Koppens06, Koppens08, Braun05}. For related reviews the reader is referred to Refs. \cite{SKhaLoss03,Zhang07,Klauser07,Coish09,Taylor07}. 
Other nanostructures in which similar situations arise are given by carbon nanotube 
quantum dots \cite{Church09}, phosphorus donors in silicon \cite{Abe04} and nitrogen vacancies in diamond.
\cite{Jel04, Child06, Hanson08}

However, the hyperfine interaction allows to access the nuclear spins efficiently. Hence, when 
it comes to utilize them instead of the electron spins for quantum information purposes, 
vice turns into virtue and the hyperfine interaction gets a very advantageous character.
Examples in this context are given by the possibility to built up an interface 
between light and nuclear spins \cite{SchCiGi08,SchCiGi09}, to polarize 
nuclear spin baths \cite{Taylor03,ChriCiGi09, ChriCiGi07}, to set up 
long-lived quantum \cite{Taylor032,Morton08} and classical \cite{Austing09} 
memory devices or to generate entanglement. \cite{ChriCiGi08}

Following the idea to take advantage of the hyperfine interaction, 
in a recent letter\cite{BJEPL} we investigated a system of two exchange coupled electron spins, 
each of which is interacting with an individual bath
of nuclear spins via the hyperfine interaction. In contrast to 
most of the approaches considered in the context of hyperfine interaction 
\cite{KhaLossGla02, KhaLossGla03, Coish04, Coish05, Coish06, Coish08}, no magnetic
field, enabling for a perturbative treatment of the problem, was applied to the electron spins.
Using exact diagonalization studies, we demonstrated that the nuclear baths can be swapped 
and fully entangled, provided they are \textit{large} enough. In order to be able to numerically consider the
required system sizes, we introduced the so-called long spin approximation (LSA). Here
we assumed homogeneous couplings within each of the baths and considered them to be highly polarized. This allows to replace them by two single long spins. Interestingly, the spectrum of the two bath model with homogeneous couplings, studied in a preceding publication \cite{ErbSchl10}, exhibits systematically degenerate multiplets under certain 
conditions. Motivated by this, we distinguished between systems with and
without inversion symmetry, i.e. a formal exchange of the central as well as the bath spins. In the latter case quantum dots of different geometry are combined to a double dot setup. Surprisingly, it turned out that here the swap performance is much better. 

In the present paper we apply the LSA in order to study the \textit{electron} spin dynamics. The results complement those of Refs. \cite{ErbSchl10, BJEPL} and in particular those of Ref. \cite{ErbSchl09}, where we studied the electron spin evolution assuming the electrons to interact with
a common bath of nuclear spins via homogeneous couplings. 

The paper is organized as follows: In Sec. \ref{model} we introduce 
the model and the methods. In particular, we in detail discuss the 
applicability of the LSA with respect to the electron spin dynamics.
We then study the spin and entanglement dynamics in the limit of an 
exchange coupling which is much larger than the hyperfine 
energy scale. Here the nuclear baths act as a perturbation. This is a particularly interesting case, as 
exceptionally long decoherence times can be expected.
In Sec. \ref{spindyn} we focus on the time evolution of the electron spins.
In a first step we study basic dynamical properties. In particular
we demonstrate that in certain parameter ranges the process of decoherence
is incomplete. Furthermore, we find a simple empirical rule describing the dynamical signatures 
of different initial states. We then quantitatively investigate the
decoherence time and the magnitude of the spin decay. As expected from Ref. \cite{ErbSchl09},
the decoherence time scales with the system size according to a power law. 
As already mentioned, in Ref. \cite{BJEPL} it was demonstrated that the nuclear spin dynamics strongly
benefits from combining quantum dots of different geometry to the double dot 
setup. In full generality, this result can be confirmed only in certain
parameter regimes. In Sec. \ref{ent} we then focus on the entanglement dynamics and demonstrate that, surprisingly,
even for a comparatively weak exchange coupling the electron spins can be fully entangled.

\begin{figure}
\begin{center}
\resizebox{0.8\linewidth}{!}{
\includegraphics{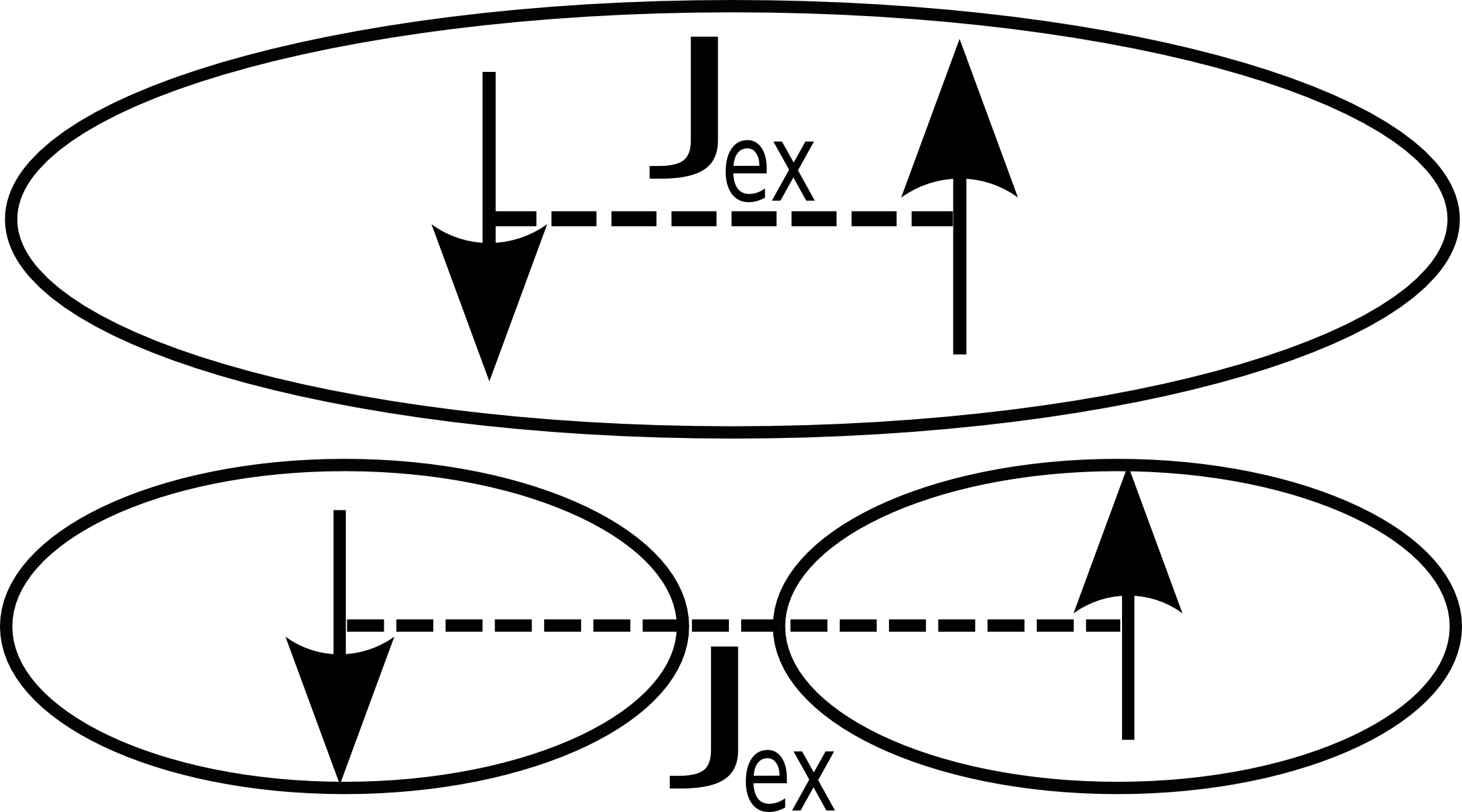}}
\end{center}
\caption{\label{Fig:baeder} 
Illustration of the one-bath and the two-bath model}
\end{figure} 

\section{Model and methods}\label{model}
The hyperfine interaction in a double quantum dot is described by the 
Hamiltonian
\begin{equation}
\label{1}
 H= \vec{S}_1 \cdot \sum_{i=1}^{N} A_i^{(1)} \vec{I}_i 
+ \vec{S}_2 \cdot \sum_{i=1}^{N} A_i^{(2)} \vec{I}_i + \Jx \vec{S}_1 \cdot \vec{S}_2 ,
\end{equation}
where $\vec{S}_i$ are the electron and $\vec{I}_i$ are the nuclear spins. 
The parameter $\Jx$ denotes an exchange coupling between the two electron 
spins, which can be adjusted in a range of $[-10^{-3}, 10^{-3}]$eV. The constants
$A_i^{(1)} $, $A_i^{(2)} $ are the hyperfine couplings of the two electron spins. 
In a realistic quantum dot, these are proportional to the electronic wave function of the $j$-th electron
at the site $\vec{r}_i$ of the $i$-th nuclear spin:
\begin{equation}
 A^{(j)}_i\propto \rvert \Psi^{(j)}(\vec{r}_i) \rvert^2
\end{equation}
For typical GaAs quantum dots this leads to an interaction with $N \sim 10^6$ nuclear spins 
and the overall hyperfine coupling strength of the $j$-th electron,
\begin{equation}
\label{Overall}
 A^{(j)}:=\sum_{i=1}^{N} A_i^{(j)},
\end{equation}
is of the order of $\left[ 10^{-4},10^{-5}\right] $eV (see Ref. \cite{SKhaLoss02}). 

Due to the spatial variation of the electronic wave function, the hyperfine 
couplings are clearly spatially dependent. However, for any set of hyperfine 
coupling constants the Hamiltonian, obviously, conserves
the total spin $\vec{J}=\vec{S}_1 + \vec{S}_2 + \sum_{i=1}^N \vec{I}_i$.
This is a very helpful symmetry for exact numerical diagonalizations
of the Hamiltonian matrix  \cite{SKhaLoss02,SKhaLoss03}, through which we
will gain the dynamics of the system in what follows. Here we consider the eigensystem of the Hamiltonian
\begin{equation}
 H\ket{\psi_i} = E_i \ket{\psi_i}
\end{equation}
and decompose the initial state $\ket{\alpha}$ into a sum of energy eigenstates:
\begin{equation}
 \ket{\alpha}=\sum_i \alpha_i \ket{\psi_i}
\end{equation}
Applying the time evolution operator $U=e^{- \frac{i}{\hbar}Ht}$ and tracing out the nuclear degrees of freedom then gives the reduced density matrix for the electrons,
\begin{eqnarray}
\label{RHO}
\nonumber  \rho_{e}(t)&=&\Tr_{n} \left( \ket{\alpha(t)}\bra{\alpha(t)} \right)\\
&=& \sum_{i,j} \alpha_i \alpha^*_j e^{-\frac{i}{\hbar}(E_i-E_j)t} \Tr_{n}\left(\ket{\psi_i}\bra{\psi_j} \right),
\end{eqnarray}
from which the dynamics of all observables can be calculated. For further details see Ref. \cite{ErbSchl09}.

There we investigated the case of two electron spins 
coupled to a common nuclear spin bath. In what follows, however, we 
consider the case of two separate baths as depicted schematically in 
Fig. \ref{Fig:baeder}. 
In the first case the two electron spins are assumed to be very close to 
each  
other so that both interact with the same group of nuclear spins, 
whereas in the present case they are spatially more separated, 
leading to an interaction with an individual group. The realistic situation 
of a double quantum dot will of course lie
between these two extreme cases.

As already mentioned, in the present paper we apply the LSA to the two
bath system. In the following subsections we give a detailed discussion
of the model with a particular focus on its limitations.

\subsection{The long spin approximation (LSA)}

Let us consider two separate 
spin baths of equal size with \textit{homogeneous} couplings
to one of the two electron spins each and introduce 
$\vec{I}_j=\sum_{i=1}^{N_j} \vec{I}_{ij}$, where the $\vec{I}_{ij}$
are the $N_j$ nuclear spins the 
$j$-th electron spin interacts with. This means that $N=N_1+N_2$ and 
$\sum_{i=1}^N \vec{I}_{i}=\vec{I}_1+\vec{I}_2$, where, for simplicity, we will consider $N_1=N_2$ in what follows. Now the squares of the total spin of
each bath are separate conserved quantities. Moreover, the same holds
for the square of any sum over a {\em subset of spins of each bath},
\begin{equation}
 \label{symm}
\left[H,\vec{I}\,^2_j \right]= \left[H, \lbrace \vec{K}_j^2 \rbrace \right]  =0\,,
\end{equation}
where we have, for the sake of brevity, 
denoted the set of all the latter  operators of the $j$-th bath 
as $\lbrace \vec{K}^2_j \rbrace$. The corresponding quantum numbers
 $\lbrace {K}_j \rbrace$ can be used to characterize specific
Clebsch-Gordan decompositions of each bath.
 
\begin{figure}
\begin{center}
\resizebox{0.8\linewidth}{!}{
\includegraphics{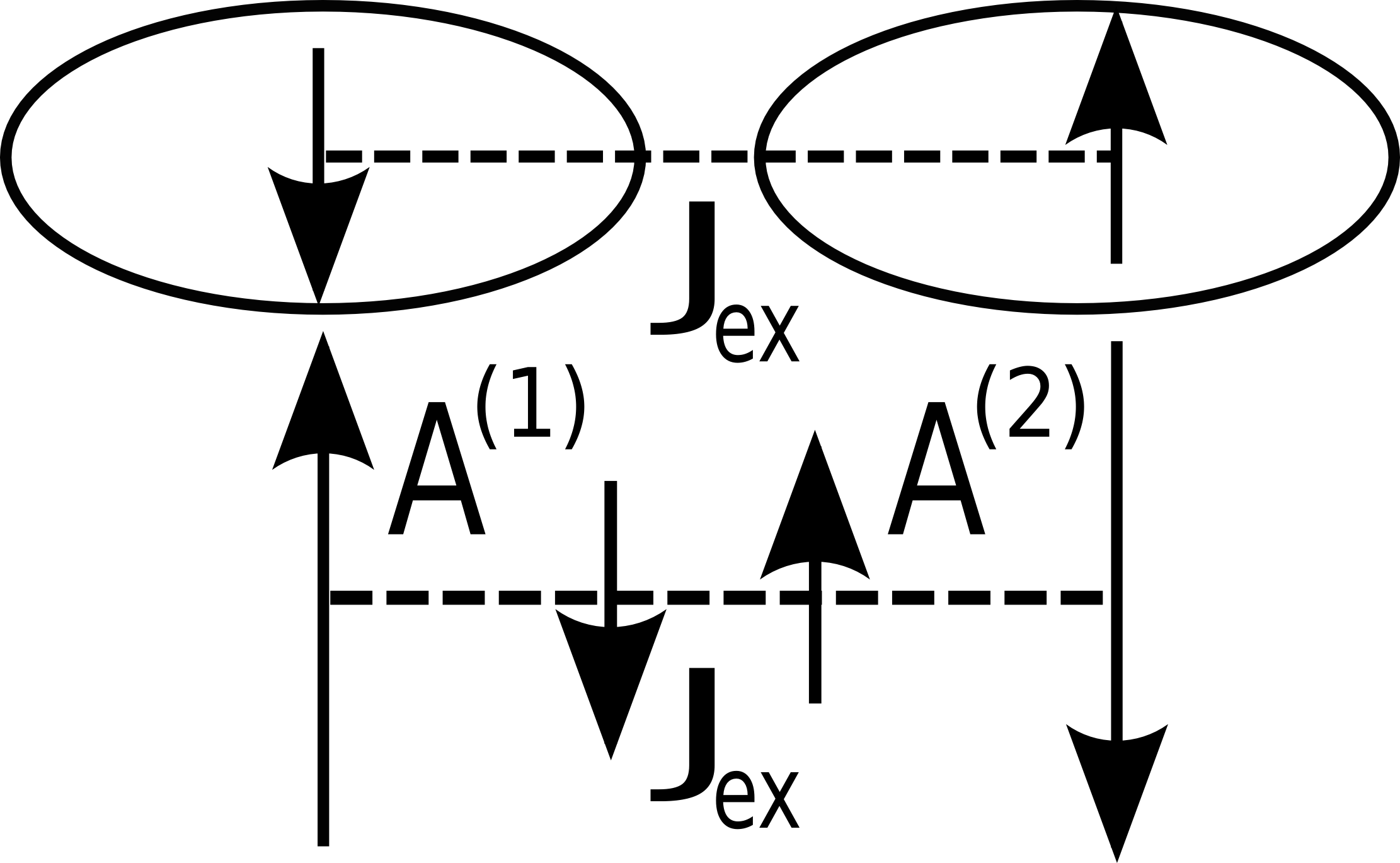}}
\end{center}
\caption{\label{Fig:longspin} The two baths are approximated by two long spins.}
\end{figure}

The initial state $\ket{\alpha}$ is given by a direct product between the initial state of the electron spins $\ket{\alpha_e}$ and the initial state of the baths $\ket{\alpha_n}$. Provided the two dots are spatially well-separated, the two resulting baths have to be considered as practically uncorrelated. Hence, the state of the nuclear baths is again a direct product between the states of the two baths, $\ket{\beta_j}$. In general such a state reads
\begin{equation}
\label{Bstate}
 \ket{\beta_j}=\sum_{I_j,m_j,\lbrace K_j \rbrace} \beta_j^{I_j,m_j ,\lbrace K_j \rbrace} \ket{I_j,m_j,\lbrace K_j \rbrace },
\end{equation}
where $\ket{I_j,m_j,\lbrace K_j\rbrace}$ are the eigenstates of $\vec{I}\,^2_j$. If the respective bath is now strongly polarized, the number of contributing multiplets in (\ref{Bstate}) drastically decreases. \cite{ErbSchl09} If we are close to full positive or negative polarization, we can drop the quantum numbers $\lbrace K_j \rbrace$ and consider the initial state to be given by $\ket{I,m_j}$ with $m_j \approx \pm I$. Due to (\ref{symm}), no ``cross terms'' between different multiplets contribute to the dynamics and all physics is then captured in the LSA Hamiltonian
\begin{equation}
\label{Ham}
 H_{\text{LSA}}=\frac{A^{(1)}}{2I} \vec{S}_1 \cdot \vec{I}_1 + \frac{A^{(2)}}{2I} \vec{S}_2 \cdot \vec{I}_2 + J_{ex} \vec{S}_1 \cdot \vec{S}_2,
\end{equation}
sketched in Fig. \ref{Fig:longspin}. The coupling constants $A^{(j)}/2I$ result directly from (\ref{Overall}) by considering $I_{ij}=1/2$: As all couplings $A_i^{(j)}$ are chosen to be equal to each other, (\ref{Overall}) yields $A^{(j)}_i=A^{(j)}/N_j$. The quantum number $I_j$ ranges from $0$ to $N_j/2$. As our model is based on highly polarized baths, we choose the maximal value. Together with $N_1=N_2$ this yields $A^j_i=A^{(j)}/2I$. 
Although the LSA Hamiltonian is not exactly solvable, the approximation of the 
baths by single long spins reduces the dimension of the problem so that 
exact numerical diagonalizations are possible on arbitrary subspaces even for 
comparatively large baths. 

Accounting for the $J^z$ symmetry, the nuclear state $\ket{\alpha_n}$ explicitly reads
\begin{equation}
\label{ini}
\ket{\alpha_n}=\ket{I,M-m-i} \ket{I,i}.
\end{equation}
Here $M$ denotes the quantum number associated with $J^z$ and $m$ the one related to $S^z$. The parameter $i$ is introduced in order to account for the deviations from $m_j= \pm I$. Hence, it has to be chosen in the vicinity of $I$ or $(M-m-I)$, respectively. Note that for an initial state which is a simple product state like (\ref{ini}), all dynamics is caused by the flip-flop terms 
\begin{equation}
\frac{1}{2} \left(S_i^+  S_j^-+S_i^-  S_j^+\right)=\vec{S}_i \cdot \vec{S}_j-S_i^3  S_j^3 
\end{equation}
in $H$ or $H_{\text{LSA}}$, respectively. This is exactly the part of the Hamiltonian, which is eliminated in most of the approaches by applying a strong magnetic field to the central spin system (see Refs. \cite{KhaLossGla02, KhaLossGla03, Coish04, Coish05, Coish06, Coish08}). In Refs. \cite{BJEPL, ErbSchl09} we also concentrated on the dynamics which are purely due to the flip-flop terms.

\subsection{Homogeneous couplings on long times scales}

In Ref.~\cite{ErbSchl09} we considered the one-bath model illustrated 
in the upper panel of Fig.~\ref{Fig:baeder} for homogeneous couplings and initial states with a very low bath polarization $p_b:=(N-2N_D)/N$ of $(1/N)$. Here $N_D$ denotes the number of flipped spins in the bath. The central spin dynamics shows periodic behaviour. This clearly has to be regarded as an artifact caused by the homogeneity of the couplings. For short time scales, meaning times much smaller than the recurrence time, the results for decoherence times found there compare well with experimental values.

As explained above, within the LSA we assume the couplings to be homogeneous and the baths to be highly polarized. 
However, high polarizations naturally lead to long time scales for the electron spin decoherence times. 
Consequently, it has to be analyzed to what extent the two assumptions of the LSA contradict each other.
In Ref. \cite{BJEPL} we already investigated this question with respect to the \textit{nuclear} spin dynamics, where we considered a Gaudin model, as corresponding to one of the first terms in (\ref{1}). We found that, qualitatively,
inhomogeneities become less important with increasing polarization. Typically, in such a context 
one would give a quantitative argument by evaluating the fidelity (to be precisely defined below) rather than studying the dynamics
on a qualitative level. However, with respect to the nuclear spin dynamics considered in Ref. \cite{BJEPL}
this does not make sense, obviously, as the bath consists of many spins so that a certain value of $\langle I^z \rangle$ 
can be realized by a whole set of nuclear states. 

In the following, we again consider a usual Gaudin model and investigate the time-averaged fidelity
$F$ with respect to homogeneous and inhomogeneous couplings via exact diagonalization. 
This is given by
\begin{equation}
F=\frac{1}{T} \int_{t=0}^T dt \lvert \braket{U_{\text{h}} \alpha}{U_{\text{ih}} \alpha} \rvert
\end{equation}
with $U_{\text{h}}$ and $U_{\text{ih}}$ being the time evolution operators for the homogeneous
or inhomogeneous Hamiltonian respectively. We choose an initial state which is a
direct product between an electron spin pointing upwards and a randomly correlated bath 
state. This is a superposition of all possible states with (in our case) real random coefficients, which
we choose in the interval $[-1,1]$. Randomly correlated states lead to highly reproducible
results and can therefore be regarded as generic. \cite{SKhaLoss02, SKhaLoss03}

\begin{figure}
\begin{center}
\resizebox{\linewidth}{!}{
\includegraphics{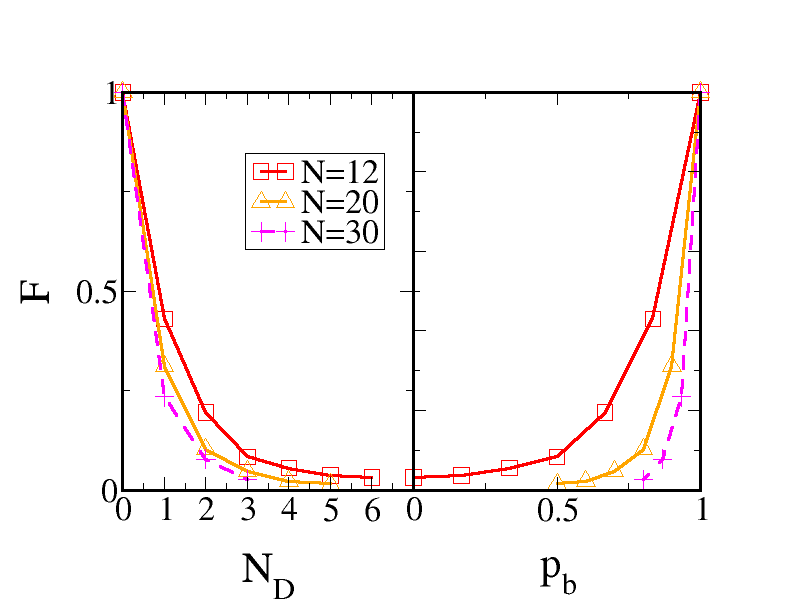}}
\end{center}
\caption{\label{Fig:01} The time-averaged fidelity is plotted against the number of flipped spins in the bath (left panel) and against the bath magnetization (right panel). We consider a randomly correlated initial bath state with coefficients in $[-1,1]$ for $N=12, 20, 30$. The fidelity strongly increases with increasing polarization. Furthermore, it decreases with an increasing number of bath spins, where the decrease gets the weaker the larger the baths become.}
\end{figure}

The results are shown in Fig. \ref{Fig:01}. We fix three different system sizes of $N=12, 20, 30$
bath spins for a reasonably long period $T=400 (\hbar/ A)$. In the left panel 
we plot $F$ against $N_D$. In order to get a better comparison between the different system sizes, in the right panel
we show the same data plotted against the bath polarization $p_b$. Obviously, the
fidelity is strongly increasing with the bath polarization. Furthermore, it decreases with an
increasing number of bath spins. Note that the different curves
approach each other with increasing number of bath spins.

\begin{figure}
\begin{center}
\resizebox{\linewidth}{!}{
\includegraphics{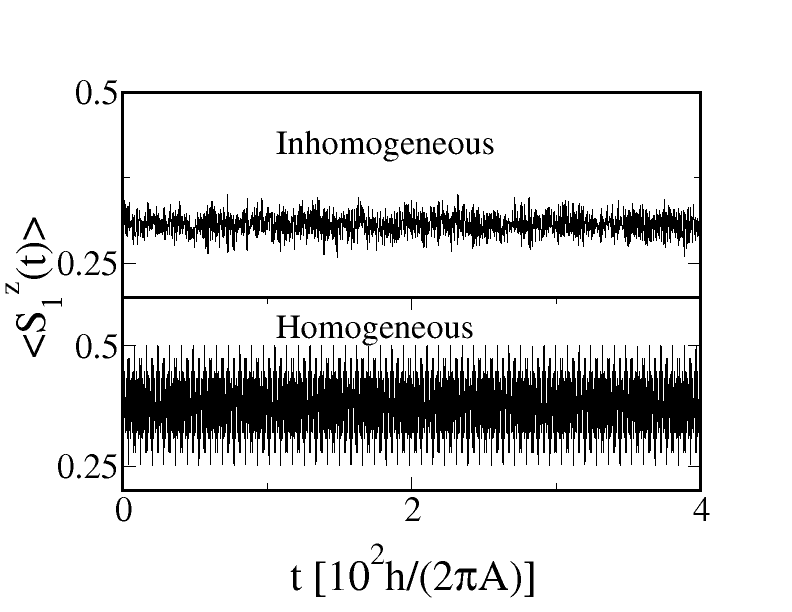}}
\end{center}
\caption{\label{Fig:02} Central spin dynamics for inhomogeneous and homogeneous couplings for $N=12$ and $N_D=4 \Leftrightarrow p_b=(1/3)$, as corresponding to $F=0.055194$. Just as in Fig. \ref{Fig:01} we consider a randomly correlated initial bath state with coefficients in $[-1,1]$. In both cases the spin is oscillating around a very similar mean value, where the amplitude is significantly larger for the case of homogeneous couplings. Hence, even in this extreme case of a very small fidelity, the dynamics are still somewhat similar to each other.}
\end{figure}

The highest experimentally feasible polarizations are around $80 \%$, as reported in Ref. \cite{Atac}. On first
sight, the results shown in Fig. \ref{Fig:01} indicate that even for such high polarizations
considering homogeneous couplings on comparatively long time scales is restricted to extremely small systems. This
would strongly contradict the purpose of the LSA. However, it
turns out that the fidelity is an \textit{extremely} sensitive measure underestimating the applicability of 
the LSA: In Fig. \ref{Fig:02} we plot the spin dynamics
for inhomogeneous and homogeneous couplings for $N_D=4 \Leftrightarrow p_b=(1/3)$, corresponding to 
a very low fidelity of $F=0.055194$. Such a small value clearly suggests that the dynamics in the inhomogeneous and the homogeneous case are fundamentally different. The amplitude of $\langle S_z(t) \rangle$ decaying to zero without any recurrence on the considered time-scales would be an example of a natural expectation for the first case (compare e.g. the results presented in Refs. \cite{SKhaLoss02, SKhaLoss03} with those of Refs. \cite{ErbSchl09, BorSt07}). Furthermore, one would guess that the time-averaged values of $\langle S_z(t)\rangle$ in the inhomogeneous and the homogeneous case strongly differ from each other. However, as can been seen from Fig. 4, neither of these expectations are met. This means that even very small fidelities correspond to a rather 
good qualitative agreement of the dynamics. Considering highly polarized baths, as done within
the LSA, it is therefore justified to choose homogeneous couplings even on comparatively long time scales. 

\section{Electron spin dynamics} \label{spindyn}

In the following we restrict ourselves to the limit $\Jx/(A/2I) \gg 1$, where we defined $A=A^{(1)} + A^{(2)}$. 
Here the dynamics is dominated by the
electron spin coupling term and the baths act as a perturbation. As a consequence, long
decoherence times, enabling e.g. to fully entangle the two electron spins, have to be expected. We will distinguish between a ``strong coupling'' and an ``ultra strong coupling'' limit. In the first case an only moderately large exchange coupling, $\Jx/A \approx 1$, is considered so that the condition $\Jx/(A/2I) \gg 1$ is realized mainly through the length of the bath spins, whereas in the second case we choose a very strong exchange coupling, meaning that here we already have $\Jx/A \gg 1$. As $I_1=I_2$, a zero ``detuning'' $\Delta := A^2-A^1$ is associated with a system invariant under inversions $1 \leftrightarrow 2$. \cite{ErbSchl10} Physically a detuning different from zero corresponds to dots of different geometry combined to a double quantum dot. Throughout the paper we will consider the dynamics on subspaces with fixed magnetization. It therefore suffices to investigate the $z$ components of the spins. Furthermore, due to their strong coupling, the dynamics of the two electron spins can be read off from each other even for $\Delta \neq 0$. Therefore, we always focus on the time evolution of the first electron spin. Note that the energy scale is given by $(A/2I)$ and consequently the time will be given in units of $hI/\pi A$. 
\begin{figure}
\begin{flushright}
\resizebox{\linewidth}{!}{
\includegraphics{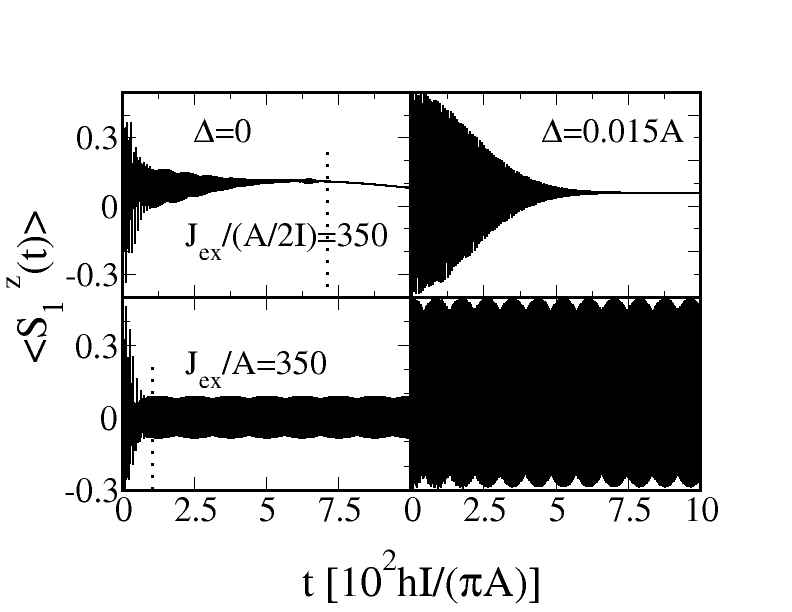}}
\end{flushright}
\caption{\label{Fig:11} Electron spin dynamics for the strong and the ultra strong coupling case $\Jx/(A/2I),\Jx/A=350$ with $\Delta=0$ in the left column and $\Delta \neq 0$ in the right one. We consider $I=120$ and $M_r=0.17$. The initial state is given by $\ket{\alpha}=\ket{\Uparrow \Downarrow}\ket{M-I,I}$. As illustrated by the dotted lines, we choose the point at which the amplitude does not change anymore as the decoherence time.
 We find very regular dynamics, where the decoherence times are obviously larger in the case of broken inversion symmetry $\Delta \neq 0$. Furthermore, the oscillations do not fully decay in the ultra strong coupling case.  }
\end{figure}

\begin{figure}
\begin{flushright}
\resizebox{\linewidth}{!}{
\includegraphics{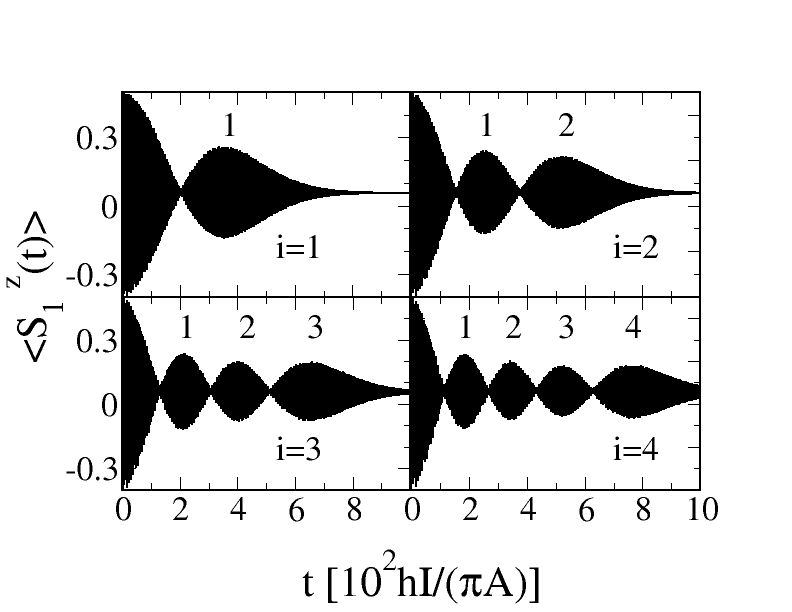}}
\end{flushright}
\caption{\label{Fig:13} Electron spin dynamics for the strong coupling case $\Jx/(A/2I)=350$ and $\Delta = 0.015$. We consider $I=120$ and $M_r=0.17$. The initial states are given by $\ket{\alpha}=\ket{\Uparrow \Downarrow}\ket{M-I+i,I-i}$ with $i=1,2,3,4$. We find dynamics with an envelope, decaying in a quite similar way to the one for $i=0$ shown in Fig. \ref{Fig:11}. However, here additional beatings occur. Their number is equal to $i$. }
\end{figure}

\subsection{Basic dynamical properties}

In order to give a basic impression of the dynamics, in Figs. \ref{Fig:11} and \ref{Fig:13} we fix $\Jx/(A/2I)=350$ for the strong and $\Jx/A=350$ for the ultra strong coupling case and plot the dynamics of the first electron spin for $I=120$. We consider the relatively low ``magnetization'' $M_r:=M/(2I+1)=0.17$. This means that we concentrate on initially nearly antiparallel baths. We study the inversion invariant case $\Delta=0$ as well as $\Delta \neq 0$. All initial states considered in Figs. \ref{Fig:11} and \ref{Fig:13} have an antiparallel electron spin configuration $\ket{\alpha_e}=\ket{\Uparrow \Downarrow}$. In Fig. \ref{Fig:11}  we consider an initial nuclear state with a maximally negative $z$ component of the first bath spin, $\vec{I}_1$. This corresponds to $i=I$ in (\ref{ini}). We clearly see that in the ultra strong coupling limit the time evolution for initial states of the above mentioned form does not fully decay. Indeed, for \textit{small, inversion invariant} systems this is also the case in the strong coupling limit. Varying $i$ slightly away from $I$, the dynamics in the ultra strong coupling case does not show any qualitative change. As can be seen in Fig. \ref{Fig:13}, this is also the case for the \textit{envelope} of the dynamics in the strong coupling limit. However, here additional beatings occur. Surprisingly, there is a clear empirical rule concerning these additional low frequency oscillations: If the $z$ component of the first bath spin deviates by $i$ from the maximal negative value, $\ket{\alpha_n}=\ket{M-I+i,I-i}$, the dynamics shows exactly $i$ beatings. In Fig. \ref{Fig:13} the case of broken inversion symmetry is considered, where the beatings are particularly pronounced. 
At the time being, we are not able to explain this effect. However, it seems that non-trivial dynamical regularities are typical for central spin models with homogeneous couplings. Indeed, in Ref. \cite{ErbSchl09} we reported on a rule for the one bath model, which relates the number of flipped spins in the initial state of the bath to the number of local extrema in the oscillations of the central spins. Also the dynamics has been calculated on a fully analytical level, we have not been able to give an explanation of these regularities. 

\subsection{Decoherence time and magnitude of the spin decay}

In direct analogy to the investigations in Ref. \cite{ErbSchl09}, in the following we investigate the scaling of the decoherence time with the spin length. It is clear that such an investigation can not yield perfectly reliable values, as the spin length is of course restricted to comparatively small values due to the limited computational power. Consider for example $M_r=0$. Here the dimension of the Hilbert space is given by $(8I+2)$, limiting the length of the spins to values of the order of $I \sim 10^2$. Still, the results give a clear idea about the type of scaling and allow for a qualitative comparison between different parameter regimes. In the following we concentrate on initial states $\ket{\alpha}=\ket{\Uparrow \Downarrow}\ket{M-I,I}$ for $M_r=0.17$. As already explained, $i$ has to be in the vicinity of $I$ and the envelope remains unaffected when varying $i$ slightly away from its maximal value. Hence, the results for $\ket{\alpha}=\ket{\Uparrow \Downarrow}\ket{M-I,I}$ can be regarded as generic. As can be seen from Fig. \ref{Fig:11}, in the ultra strong coupling limit $\langle S_1^z(t) \rangle$ does not decay to a constant value, but oscillations of quite regular shape remain, i.e. the decoherence process is not complete. Therefore we define the time from which on the amplitude does not change anymore as the decoherence time. Numerically this is realized by dividing the time axis in intervals with a length larger than the period of the regular oscillations and determining the maximal value in each interval. If this value does not change anymore over a fixed number of intervals, the lower bound of the first interval in which the respective value appeared is chosen as the decoherence time. In the left panels of Fig. \ref{Fig:11} this choice is illustrated by the dotted lines.
\begin{figure}
\begin{flushright}
\resizebox{\linewidth}{!}{
\includegraphics{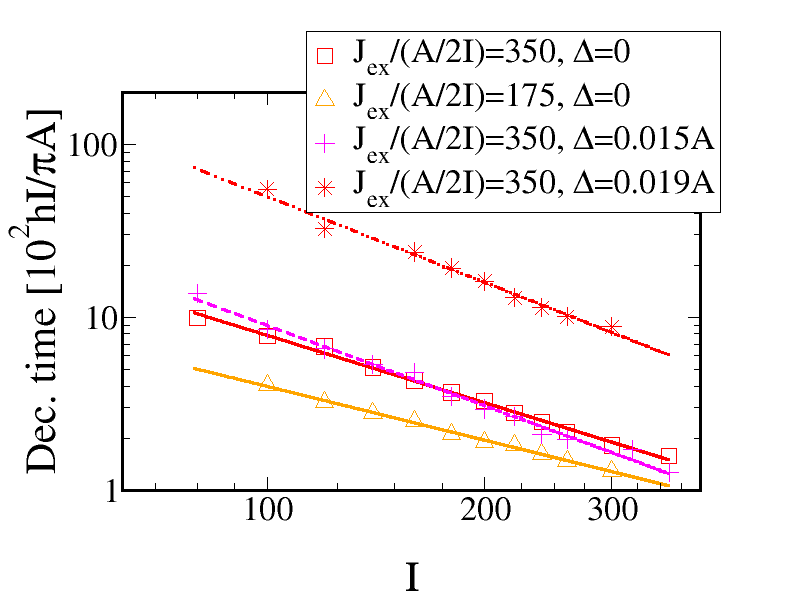}}
\end{flushright}
\caption{\label{Fig:21} Scaling of the decoherence time with the bath spin length $I$ on a double logarithmic scale for the strong coupling case. Different coupling ratios $\Jx/(A/2I)$ and detunings $\Delta$ are considered. We choose the initial state $\ket{\alpha}=\ket{\Uparrow \Downarrow}\ket{M-I,I}$ for $M_r = 0.17$. We find power laws $\sim I^{- \nu}$ with $\nu=1.29$ ($\Jx/(A/2I)=350, \Delta=0$), $\nu=1.03$ ($\Jx/(A/2I)=175, \Delta=0$), $\nu=1.54$ ($\Jx/(A/2I)=350, \Delta=0.015A$), and $\nu=1.65$ ($\Jx /(A/2I)=350, \Delta=0.019A$). Although  the limit $\Jx/(A/2I) \gg 1$ is left unaltered, surprisingly the scaling changes with the coupling ratio. Breaking the inversion symmetry leads to an increase of the parameter $\nu$.  }
\end{figure}
\begin{figure}
\begin{flushright}
\resizebox{\linewidth}{!}{
\includegraphics{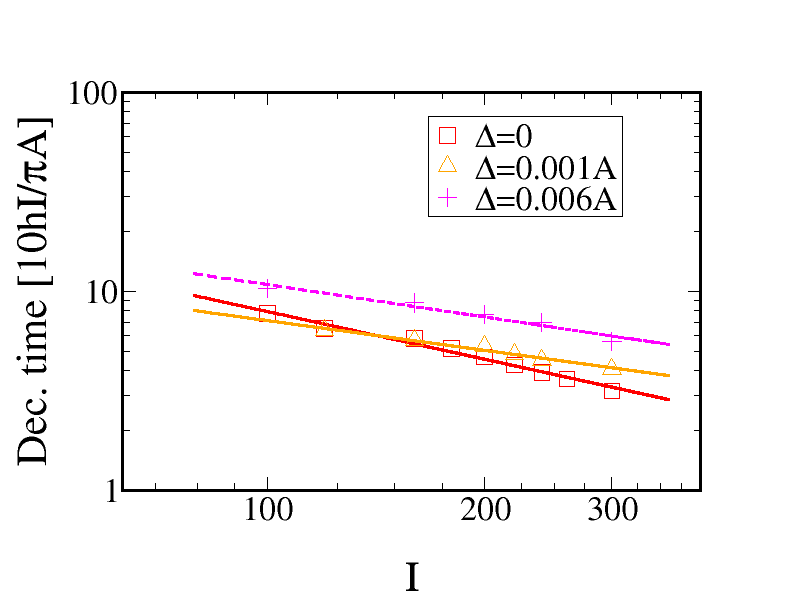}}
\end{flushright}
\caption{\label{Fig:22} Scaling of the decoherence time with the bath spin length $I$ on a double logarithmic scale for the ultra strong coupling case. We fix $\Jx/A=350$ and vary $\Delta$ weakly away from zero. We choose the initial state $\ket{\alpha}=\ket{\Uparrow \Downarrow}\ket{M-I,I}$ for $M_r = 0.17$. The curves are again fitted to power laws $\sim I^{- \nu}$ with $\nu=0.79$ ($\Delta=0$), $\nu=0.49$ ($\Delta=0.001A$), and $\nu=0.54$ ($\Delta=0.006A$). Note that the absolute of the decoherence times are smaller than in the strong coupling case because the dynamics does not fully decay. Breaking the inversion symmetry leads to a decrease of $\nu$ to $\nu \approx 0.5$, which is the value found in Ref \cite{ErbSchl09} for the one bath model. }
\end{figure}

In Ref. \cite{ErbSchl09} it has been shown for the one bath model that the decoherence time scales with the size of the bath according to a power law $ \sim N^{-\nu}$. Indeed, we find the same behavior for the present case. In Figs. \ref{Fig:21}, \ref{Fig:22} the decoherence times for the strong and the ultra strong coupling case are plotted against the spin length $I$ on a double logarithmic scale. For the inversion symmetric case we consider $\Jx /(A/2I), \Jx/A= 350$ and $\Jx/(A/2I)=175$, where it obviously does not make any sense to choose a second value for the ultra strong coupling limit. For the case of broken inversion symmetry we fix $\Jx/(A/2I), \Jx/A= 350$ and fix $\Delta=0.015A, 0.019A$ for the strong coupling and $\Delta=0.001A, 0.006A$ for the ultra strong coupling case. The values for the latter are chosen to be particularly small, because, as exemplified in the bottom panels of Fig. \ref{Fig:11}, the dynamics is highly sensitive with respect to a change of the detuning and becomes completely coherent on any relevant time scale for larger values.

As expected, for the strong coupling case the decoherence time is scaling much stronger than for the ultra strong coupling limit. Note that the values for the ultra strong case are much smaller only due to the fact that here the dynamics does not fully decay. As can be seen from Fig. \ref{Fig:21}, in the strong coupling limit the scaling does change significantly with the coupling ratio $\Jx/(A/2I)$. This is surprising as a small change in the ratio leaves the limit $\Jx /(A/2I) \gg 1$ unaltered and hence one would expect the scaling to be insensitive against a change of the coupling ratio. Furthermore, the absolute values of the decoherence time clearly decrease with decreasing coupling ratio $\Jx/(A/2I)$ as expected. However, as a counterintuitive effect the scaling with the system size turns out to be weaker for the smaller of the two ratios. Breaking the inversion symmetry has a significant effect in the strong as well as the ultra strong coupling limit. In the first case the exponent $\nu$ increases, whereas in the latter it decreases to $\nu \approx 0.5$. This is the value derived in Ref. \cite{ErbSchl09} for the one bath model.
\begin{figure}
\begin{flushright}
\resizebox{\linewidth}{!}{
\includegraphics{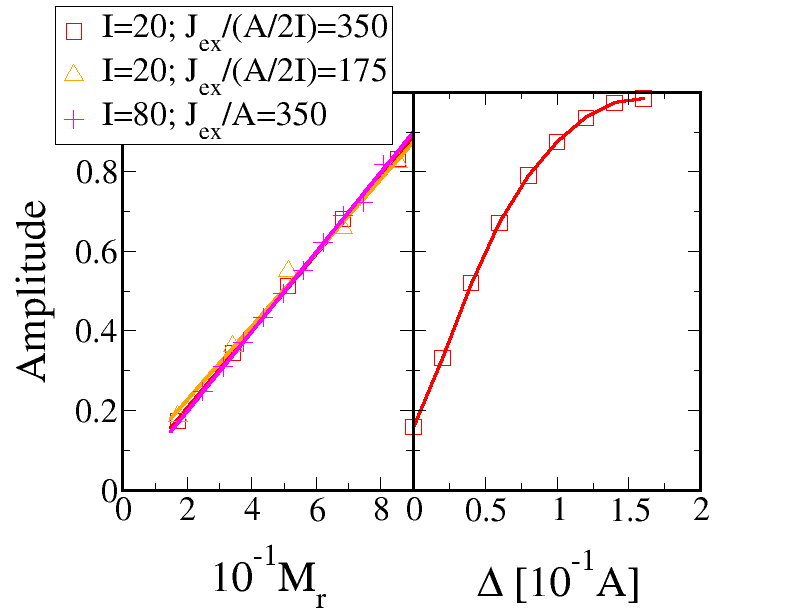}}
\end{flushright}
\caption{\label{Fig:3} Scaling of the amplitude, given by the difference of a local maximum and the following local minimum, with the magnetization and the detuning for the cases, where the dynamics does not fully decay. In the left panel we fix $\Delta=0$. For the strong coupling case we consider $I=20$ and $\Jx/(A/2I)=350, 175$. For the ultra strong coupling case we fix $I=80$ and $\Jx/A=350$. We choose the initial state $\ket{\alpha}=\ket{\Uparrow \Downarrow}\ket{M-I,I}$. In all cases we find a linear dependence of the amplitude on the magnetization with gradients $a$ close to one. The values are given $a=0.97$ ($\Jx/(A/2I)=350$), $a=0.92$ ($\Jx/(A/2I)=175$), and $a=1.0$ ($\Jx/A=350$). In the right panel we plot the the amplitude against the detuning for $M_r\approx 0.17$. Here we find a highly non-linear dependence.}
\end{figure}

As explained in the preceding subsection, in the ultra strong coupling case the dynamics does not show full decoherence. If the spin length $I$ is small and we have $\Delta=0$ this is also the case for only strongly coupled electron spins. We now analyze the scaling of the decaying part of the dynamics as a function of the magnetization and, in the ultra strong coupling case, as a function of the detuning $\Delta$. Concerning the strong coupling case the results are, obviously, only of fundamental interest. Even in SiGe and carbon based quantum dots (only around $4.7 \%$ of the Si isotopes are spin carrying\cite{Bou}, in carbon even only $1\%$\cite{Fischer}), the electron spins interact with a few thousands of nuclear spins.

Note that our LSA model is valid only for relatively small and relatively large magnetizations $M_r$, which corresponds to nuclear spin baths highly polarized in either the same or opposite directions. However, in Fig. \ref{Fig:3} we plot the amplitude, defined as the difference between a local maximum and the following local minimum, for the whole range of $M_r$, which corresponds to either parallel or anti. We consider the strong as well as the ultra strong coupling case. For the first one we fix $I=20$ and two values for the coupling ratio $\Jx /(A/2I)=350,175$. In both cases we find a linear dependence with a gradient close to one, meaning that the ratio does not significantly influence the decaying part. Concerning the ultra strong coupling case, we set $I=80$ and consider $\Jx/A=350$. The scaling is practically identical to the one for the strong coupling case. As already discussed in the preceding section, we found that for the ultra strong coupling case the decaying part is not only influenced by the magnetization but also by the detuning. In the right panel of Fig. \ref{Fig:3} we plot the amplitude against the detuning for a fixed magnetization $M_r=0.17$. In contrast to a variation of the magnetization, here we find a highly non-linear dependence, which can not be fitted by some simple power law.

In Ref. \cite{BJEPL} we demonstrated that a non-zero detuning is very advantageous with respect to swapping and entangling the nuclear spin baths. When it comes to the electron spin dynamics, however, in general this is the case only the ultra strong coupling limit. 

\section{Entanglement dynamics}\label{ent}

We now close the discussion of the electron spin dynamics with an investigation of the entanglement between the two electron spins. In order to quantify the non-classical correlations, we consider the concurrence defined by \cite{Wootters97}
\begin{equation}
C(t)=\text{max}\left\lbrace 0, \sqrt{\lambda_1}-\sqrt{\lambda_2}-\sqrt{\lambda_3}-\sqrt{\lambda_4} \right\rbrace ,
\end{equation}
where $\lambda_i$ are the eigenvalues of the non-hermitian matrix $\rho_{e}(t) \tilde{\rho}_e(t)$ in decreasing order. Here $\tilde{\rho}_e(t)$ is given by $\left(\sigma_y \otimes \sigma_y \right) \rho^*_e(t) \left(\sigma_y \otimes \sigma_y \right)$, where $\rho^*_e(t)$ denotes the complex conjugate of $\rho_e(t)$-the reduced density matrix of the electrons as defined in (\ref{RHO}).

\begin{figure}
\begin{flushright}
\resizebox{\linewidth}{!}{
\includegraphics{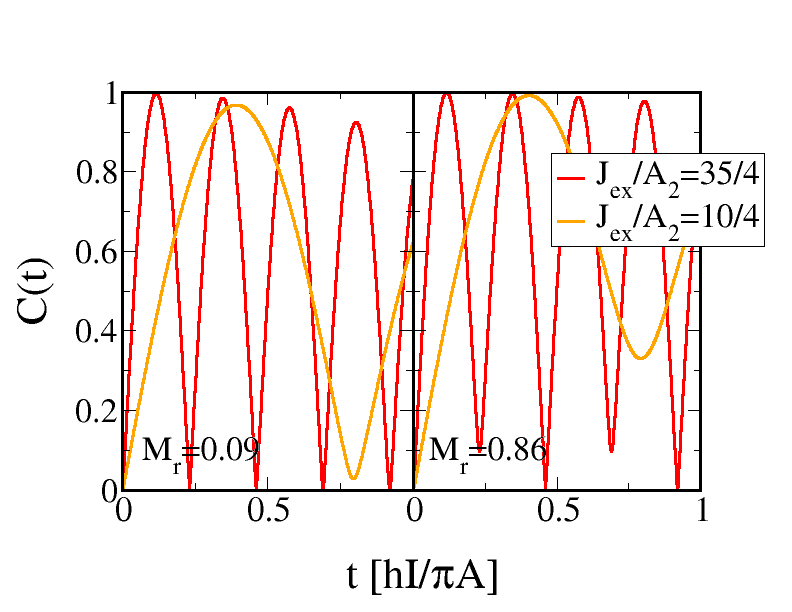}}
\end{flushright}
\caption{\label{Fig:4} Time evolution of the concurrence between the two central spins for $I=40$. We consider the case of a low polarization ($M_r=0.09$) in the left panel and the case of a high polarization ($M_r=0.86$) in the right panel. In both cases we choose a detuning of $\Delta=0.019A$ and consider $\Jx/A_2=35/4$, which satisfies (\ref{Cond}), and $\Jx/A_2=10/4$, which violates (\ref{Cond}). The nuclear spins are in a randomly correlated state with coefficients in $[-1,1]$ initially. One clearly sees that if (\ref{Cond}) is satisfied, the concurrence becomes one, whereas for a stronger coupling to the baths the electron spins can not be fully entangled.}
\end{figure}

In the following, we ask to what extent it is possible to entangle initially uncorrelated electron spins. Therefore, we again consider initial states with electron spin configurations $\ket{\Uparrow \Downarrow}$. In particular we are interested in a lower bound for the ratio $\Jx/(A/2I)$, meaning that we adjust the couplings $\Jx,A_1,A_2$ to the lowest possible ratio so that the concurrence still becomes equal to one. As to be expected, the lower bound lies in the ultra strong coupling limit. However, surprisingly it turns out that it is not determined by the ratio $\Jx/A$, but only by $\Jx/\text{max}\lbrace A_1,A_2 \rbrace$. The concrete value of this ratio depends on the initial state of the nuclear spins. An upper bound is given by the (,as explained above, unphysical) case of randomly correlated states. As an empirical rule of thumb here we find:
\begin{equation}
\label{Cond}
 \frac{\Jx}{\text{max}\lbrace A_1,A_2\rbrace}\geq 8.8
\end{equation}
In Fig. \ref{Fig:4} we illustrate the rule by plotting the dynamics for randomly correlated initial states with coefficients in $[-1,1]$ by considering parameters satisfying and violating (\ref{Cond}). We choose a rather small system of $I=40$ and concentrate on the case of broken inversion symmetry $\Delta = 0.019A$. We plot the time evolution for a low polarization of $M_r=0.09$ in the left panel and fix a rather high polarization of $M_r=0.86$ in the right panel. It is visible that the maximal value of the entanglement drops slightly under one if (\ref{Cond}) is violated.

\section{Conclusion} \label{conc}
In summary, we numerically studied the electron spin and entanglement dynamics in a system of two strongly coupled electron spins, each of which is interacting with an individual bath of nuclear spins via the hyperfine interaction. We applied the LSA introduced in Ref. \cite{BJEPL}, where the two baths are replaced by two single long spins, and focused on the limit of an exchange coupling much larger than the hyperfine energy scale. Here we distinguished between a strong and an ultra strong coupling case. We demonstrated that the decoherence time scales with the size of the baths according to a power law. As expected, it turned out that the decaying part decreases with increasing polarization. However, surprisingly it also decreases with increasing detuning, provided the electrons are bound ultra strongly. Hence, with respect to the electron spin dynamics the advantageous character of a non-zero detuning, found in Ref. \cite{BJEPL} for the time evolution of the nuclear baths, can only be confirmed in the ultra strong coupling limit. Finally, we demonstrated that it is possible to fully entangle the electron spins even for a comparatively weak exchange coupling.

\section{Acknowledgements}
This work was supported by DFG via SFB 631. 

\end{document}